\documentclass[final,5p,times,twocolumn,showpacs,superscriptaddress]{elsarticle}
\makeatletter
\def\ps@pprintTitle{%
   \let\@oddhead\@empty
   \let\@evenhead\@empty
   \let\@oddfoot\@empty
   \let\@evenfoot\@oddfoot
}
\makeatother
\usepackage{amssymb}
\usepackage{graphicx}
\usepackage{dcolumn}
\usepackage{amsmath}
\usepackage{bm}
\usepackage{epsfig}
\usepackage{hyperref}

\journal{Physica A}

\allowdisplaybreaks

\def\defi{{\buildrel \;def\; \over =}}
\newcommand{\be}{\begin{equation}}
\newcommand{\ee}{\end{equation}}

\newcommand{\mediaT}[1]{\left\langle #1 \right\rangle}

\newcommand{\media}[1]{\langle #1 \rangle}

\begin{document}


\title{1D Three-state mean-field Potts model
with first- and second-order phase transitions}

\author[UFBA]{Massimo Ostilli}
\author[UAEU]{Farrukh Mukhamedov}

\address[UFBA]{Instituto de
    F\'isica, Universidade Federal da Bahia, Salvador, Brazil}
\address[UAEU]{Department of Mathematical Sciences, College of
  Science, United Arab Emirates University, Al Ain, Abu Dhabi, UAE}

\begin{abstract}
We analyze a three-state Potts model built over a lattice ring, with coupling $J_0$,
and the fully connected graph, with coupling $J$.
This model is effectively mean-field and can be exactly solved
by using transfer-matrix method and Cardano formula.
When $J$ and $J_0$ are both ferromagnetic, the model has a first-order phase
transition which turns out to be a smooth modification of the known
phase transition of the traditional mean-field Potts model ($J_0=0$),
despite, as we prove, the connected correlation functions are now non zero, even in the paramagnetic phase.
Furthermore, besides the first-order transition,
there exists also a hidden continuous transition at a temperature below which the symmetric metastable state
ceases to exist.
When $J$ is ferromagnetic and $J_0$ antiferromagnetic, a similar
antiferromagnetic counterpart phase transition scenario applies. Quite interestingly, differently from the
Ising-like two-state case, for large values of the antiferromagnetic coupling $J_0$, the critical temperature
of the system tends to a finite value. Similarly, also the latent heat per spin tends to a finite constant in the
limit of $J_0\to -\infty$.
\end{abstract}

\maketitle

  \begin{keyword}
Exact Results \sep Potts Model \sep Phase Transitions \sep Effective Mean Field
  \end{keyword}

\section{Introduction}
The mean-field concept is a fundamental
paradigm in theoretical physics and its interdisciplinary applications. It consists
in replacing the interactions acting on a particle with an effective
external field to be determined self-consistently.
The power of this approach manifests in two ways:
on one hand, it allows to face analytically, in a first approximation, any given model;
on the other hand, it provides a powerful understanding of the physics
of the model. In fact, even though very approximate, the mean-field solution is often
pedagogically deeper than the understanding one would get
from a possible exact solution (if any).
In particular, it would be harder to understand
the concept of the collective behavior and the phase transitions
of a system without a suitable mean-field theory.

At the mathematical base of the mean-field theory there are models
which are exactly solvable by a mean-field technique:
the mean-field models.
These models represent the limit cases of more realistic models
in which one or more parameters are typically send to 0 or to $\infty$
so that the mean-field approximation becomes exact.
Traditionally, the concept of the mean-field models is associated with
the absence of correlations in the thermodynamic limit.
In \cite{MF} (see also \cite{Bogoliubov} and \cite{Perk}) we have shown that this condition is only a sufficient condition
for the system to be mean-field, but in general it is not necessary.
There exist in fact infinite many models having both non zero correlations
and a mean-field character. For example, if $H_0$ is an arbitrary Hamiltonian,
the model $H=H_0+\Delta H$, with $\Delta H$ a general fully connected
interaction, is mean-field, in the sense that we can exactly replace
the interactions acting on a particle with an effective external
field to be determined self-consistently. However, now, the presence
of the term $H_0$ gives rise to non zero correlations whenever $H_0$
has short-range interactions \footnote{The case of power-law like long-range interactions is more subtle.
See the Conclusions in \cite{MF}.}.

Of course, unlike the traditional mean-field models
(where $H_0=0$ and there are not short-range correlations),
the arbitrariness of $H_0$ lets it open now a very richer
scenario of phase transitions.
In particular, it can be shown that,
when $H_0$ has antiferromagnetic interactions, inversion transition phenomena
and first-order phase transitions may set in \cite{SW}.
More in general, the phase transition scenario associated
to the term $\Delta H$ can change drastically when
$H_0$ has antiferromagnetic couplings.

In recent years, a renewed attention toward models having both short- and long-range
interactions, has been drawn due to the
importance of small-world networks \cite{Watts}, where a finite-dimensional
and an infinite-dimensional character are both present in the network structure.
As expected, such models turn out to be mean-field, at least for what concerns their critical behavior.
However, rather than
a theorem, except for the Ising case near the critical point \cite{Hastings2,SW},
and a few examples in one dimension \cite{Skantzos,SkantzosCav}, this turns out to be an empirical fact.
An exact analytically treatment, even not rigorous and confined to relatively simple models, is still far
from being reached when short-range correlations are present, as happens in a small-world network.
On the other hand, the models introduced in \cite{MF} can be seen as ideal small-world networks in which the random
connectivity of the graph goes to the system size $N$ and the coupling $J$ associated to the long-range interactions is
replaced by $J/N$. Clearly, without a serious understanding of the more basic
models presented in \cite{MF}, the analytical study of the small-world networks and its generalizations (including
the scale-free case \cite{DMAB}, which for the Ising case has been analyzed in \cite{SWSF}), will remain impossible.

In this spirit, in the present paper we analyze a simple and yet rich model:
a case in which $H_0$ is a one-dimensional three-state
Potts model \cite{Wu} and $\Delta H$ is the traditional three-state mean-field term,
\textit{i.e.}, the ordinary fully-connected interaction.
The mean-field equations in this case are sufficiently simple
to be exactly solved via the transfer matrix method and the Cardano formula for cubic equations.
As expected, similarly to the analog Ising case \cite{SW},
the presence of a non zero ferromagnetic coupling, $J_0>0$, in $H_0$,
alters only smoothly the phase diagram of the system
characterized by a first-order phase transition.
The difference with respect to the case without $H_0$ is that, for $H_0\neq 0$,
the connected correlations functions are now not zero.
Besides the first-order transition,
there emerges also a second-order transition.
This continuous transition is not stable
(the corresponding free energy being not a local minimum but a saddle point),
however it corresponds to a non trivial solution of the mean-field equations
and occurs at a temperature $T_c^{(SO)}<T_c^{(\mathrm{FO})}$ below which the symmetric solution ceases to exist as a metastable state.
When $H_0$ has an antiferromagnetic coupling, $J_0<0$,
a similar phase transition scenario still applies but characterized
by an antiferromagnetic order and, quite interestingly, differently from the
Ising-like two-state case, in the limit $J_0\to -\infty$, $T_c^{(\mathrm{FO})}$
tends to a finite value. Moreover, we show that in the same limit also the latent heat per spin tends to a finite constant.
Finally, we prove that the connected correlation functions are not zero and evaluate them in
a specific case.

\section{Generalized mean-field Potts models}
In the spirit of \cite{MF}, we introduce now a model built by using both
finite-dimensional and infinite dimensional Hamiltonian terms.
A generalized mean-field Potts model, \textit{i.e.}, a model where each variable $\sigma$
can take $q$ values, $\sigma=1,\ldots,q$, can be defined through the following Hamiltonian
\begin{eqnarray}
\label{HgPotts}
H=H_0(\{\sigma_i\})-\frac{J}{N}\sum_{i<j}\delta(\sigma_i,\sigma_j),
\end{eqnarray}
where $\delta(\sigma,\sigma')$ is the Kronecker delta function
and $H_0$ is any $q$-states Potts Hamiltonian with no external field.
Let us rewrite $H$ as (up to terms negligible for $N\to\infty$)
\begin{eqnarray}
\label{HgPotts1}
H=H_0(\{\sigma_i\})-\frac{J}{N}\sum_\sigma \left[\sum_{i}\delta(\sigma_i,\sigma)\right]^2.
\end{eqnarray}
As done in \cite{MF}, from Eq. (\ref{HgPotts1}) we see that, by introducing $q$ independent Gaussian variables $x_\sigma$, we
can evaluate the partition function, $Z= \sum_{\{\sigma_i\}}\exp(-\beta H(\{\sigma_i\}))$, as
\begin{eqnarray}
\label{Z1Potts}
Z\propto \int \prod_{\sigma=1}^{q} d x_\sigma ~ e^{-N\left[
\sum_\sigma \frac{\beta J x_\sigma^2}{2}+
\beta f_0(\beta J x_{1},\ldots, \beta J x_{q})
\right]},
\end{eqnarray}
where $f_0(\beta h_{1},\ldots, \beta h_{q})$ is the free energy density of the
Potts model governed by $H_0$ at the temperature $1/\beta$ and in the presence of a $q$-component external field
$\bm{h}\defi(h_{1},\ldots,h_{q})$ via $H_0\to H_0-\sum_\sigma h_\sigma \sum_i\delta(\sigma_i,\sigma)$.
By using the saddle point method, from Eq. (\ref{Z1Potts}) we find that, if $x_{0;\sigma}(\beta h_{1},\ldots \beta h_{q})$
is the order parameter for $H_0$ as a function of a $q$-component external field~\footnote{
We suppose, for simplicity, that the order parameter associated to $H_0$ (the pure model),
does not depend on the vertex position.
},
$\media{\delta(\sigma_i,\sigma)}_0=x_{0;\sigma}(\beta h_{1},\ldots \beta h_{q})$, then,
in the thermodynamic limit,
the order parameter for $H$, $x_{\sigma}=\media{\delta(\sigma_i,\sigma)}$, satisfies the system
\begin{eqnarray}
\label{Potts}
x_{\sigma}=x_{0;\sigma}\left(\beta J x_{1},\ldots, \beta J x_{q}\right), \quad \sigma=1,\ldots,q
\end{eqnarray}
and the free energy $f$ is given by
\begin{eqnarray}
\label{Pottsf}
\beta f= \sum_\sigma \frac{\beta J x_\sigma^2}{2}+
\beta f_0(\beta J x_{1},\ldots, \beta J x_{q}).
\end{eqnarray}

When $J<0$, the approach with the Gaussian variables is not valid since
the Gaussian integral diverges. Yet, the saddle point Eqs. (\ref{Potts})
are still exact, as derived from the general theorem presented in \cite{MF}
(while the free energy has a different form with respect to Eq. (\ref{Pottsf})).

Concerning the connected correlation function $C$, as a general rule we have \cite{MF} 
\begin{eqnarray}
\label{Corr}
C=C_0(\beta J x_{1},\ldots, \beta J x_{q}) +\mathrm{finite~size~effects},
\end{eqnarray}
where $C_0(\beta h_{1},\ldots,\beta h_{q})$ is the connected correlation function of the 
Potts model governed by $H_0$ at the temperature $1/\beta$ and in the presence of a $q$-component external field
$\bm{h}\defi(h_{1},\ldots,h_{q})$.


\section{The traditional mean-field Potts model $(H_0=0)$}
Before facing the analysis of our model, we want to briefly recall
the traditional mean-field Potts model defined as in Eq. (\ref{HgPotts}) with $H_0=0$.
\subsection{The pure model}
The use of Eqs. (\ref{Potts})-(\ref{Pottsf}) in this case may seem not necessary but it is instructive.
To apply Eqs. (\ref{Potts})-(\ref{Pottsf}) to the present case, we need to solve the corresponding pure model, which is
a Potts model without interaction but in the presence of a uniform external field, $\bm{h}$.
We have therefore to calculate the
following trivial partition function, $Z_0(\bm{h})$, which differs from $Z$
for the absence of the fully-connected (long-range)
interaction:
\begin{eqnarray}
\label{Z0}
Z_0(\bm{h})=\sum_{\sigma_1,\ldots,\sigma_N}e^{\beta\sum_{\sigma} h_\sigma N_\sigma},
\end{eqnarray}
where $N_\sigma=\sum_i\delta(\sigma,\sigma_i)$.
We have
\begin{eqnarray}
\label{Potts0}
x_{0;\sigma}\left(\beta h_{1},\ldots, \beta h_{q}\right)=\frac{e^{\beta h_\sigma}}{\sum_{\sigma'}e^{\beta h_{\sigma'}}},
\end{eqnarray}
\begin{eqnarray}
\label{Pottsf0}
\beta f_0(\beta h_{1},\ldots, \beta h_{q})=-\log\left(\sum_\sigma e^{\beta h_{\sigma}}\right).
\end{eqnarray}

\subsection{The mean-field model}
\begin{figure}
\includegraphics[scale=0.35]{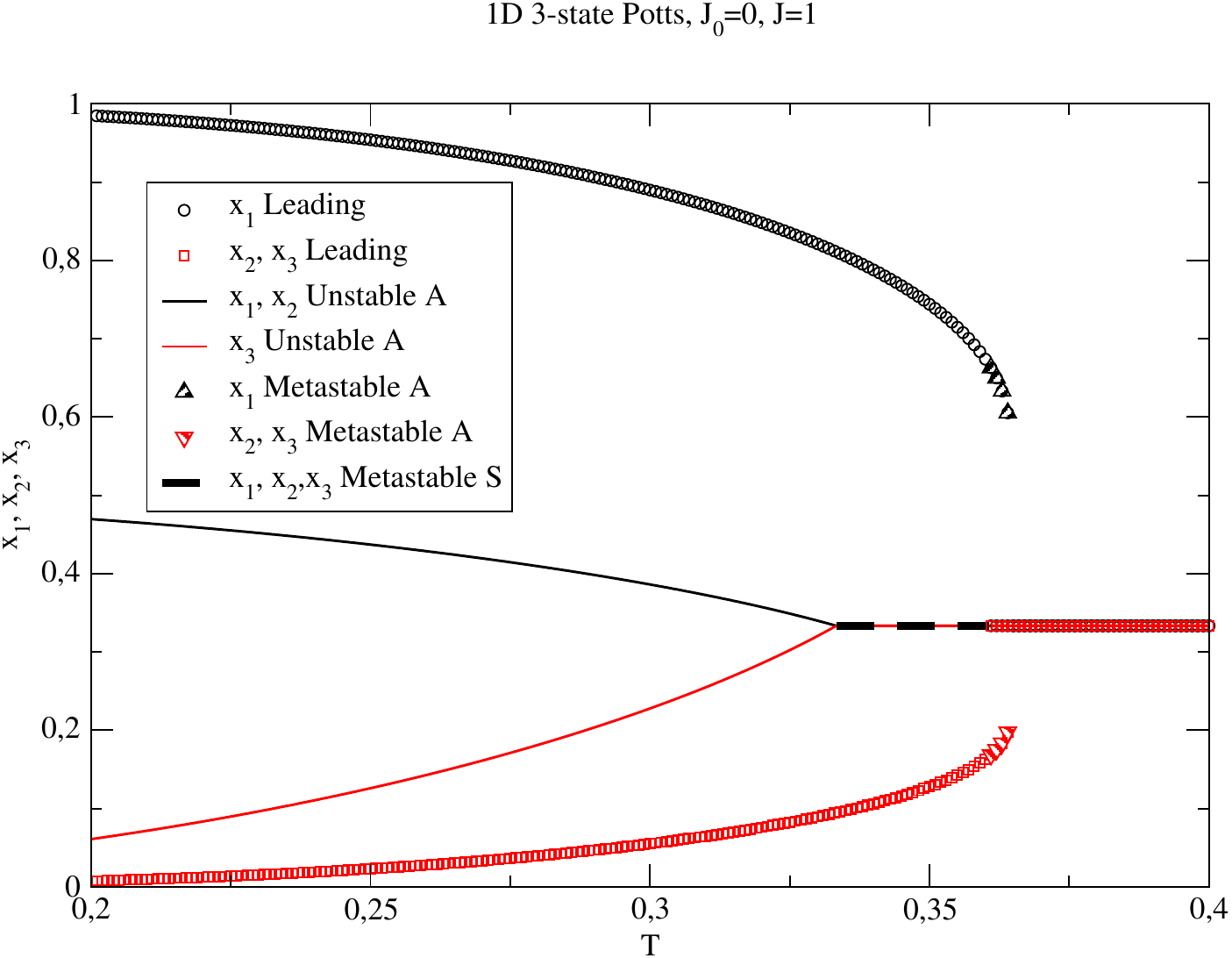}
\caption{(Color online) Magnetizations for the case $J=1$ and $J_0=0$ (equivalent to the traditional mean-field Potts model).
In the figure: ``Leading'' stands for the thermodynamic stable state having the lower
free-energy, ``Metastable A'' and ``Metastable S'' stand for the thermodynamic stable states having the higher
free-energy, while ``Unstable S'' and ``Unstable A'' stand for the thermodynamic
unstable states; A or S stand for asymmetrical or symmetrical, respectively.
For any $J_0\geq 0$ (which includes the present case $J_0=0$), the Leading states lie on the subspaces $x_{i_1}>x_{i_2}=x_{i_3}$
(in this figure only the case $x_1>x_2=x_3$ is shown),
while the Unstable A states lie on the subspaces $x_{i_1}=x_{i_2}>x_{i_3}$, where
$i_1,i_2,i_3$, is any permutation of the set of indices $\left\{1,2,3\right\}$
and $x_{i_1}+x_{i_2}+x_{i_3}=1$ (in this figure only the case $x_1=x_2>x_3$ is shown).
Note that the asymptotic values of the magnetizations toward $T=0$
are $x_{i_1}=1,~x_{i_2}=x_{i_3}=0$ and $x_{i_1}=x_{i_2}=1/2,~x_{i_3}=0$, for the Leading
and Unstable A states, respectively. The first-order phase transition occurs at $T_c^{(\mathrm{FO})}=0.3607$, while the second-order one
at $T_c^{(\mathrm{SO})}=1/3$. The Metastable A state begins at $T=0.364$ and ends at $T_c^{(\mathrm{FO})}$.
}
\label{fig1}
\end{figure}

\begin{figure}
\includegraphics[scale=0.35]{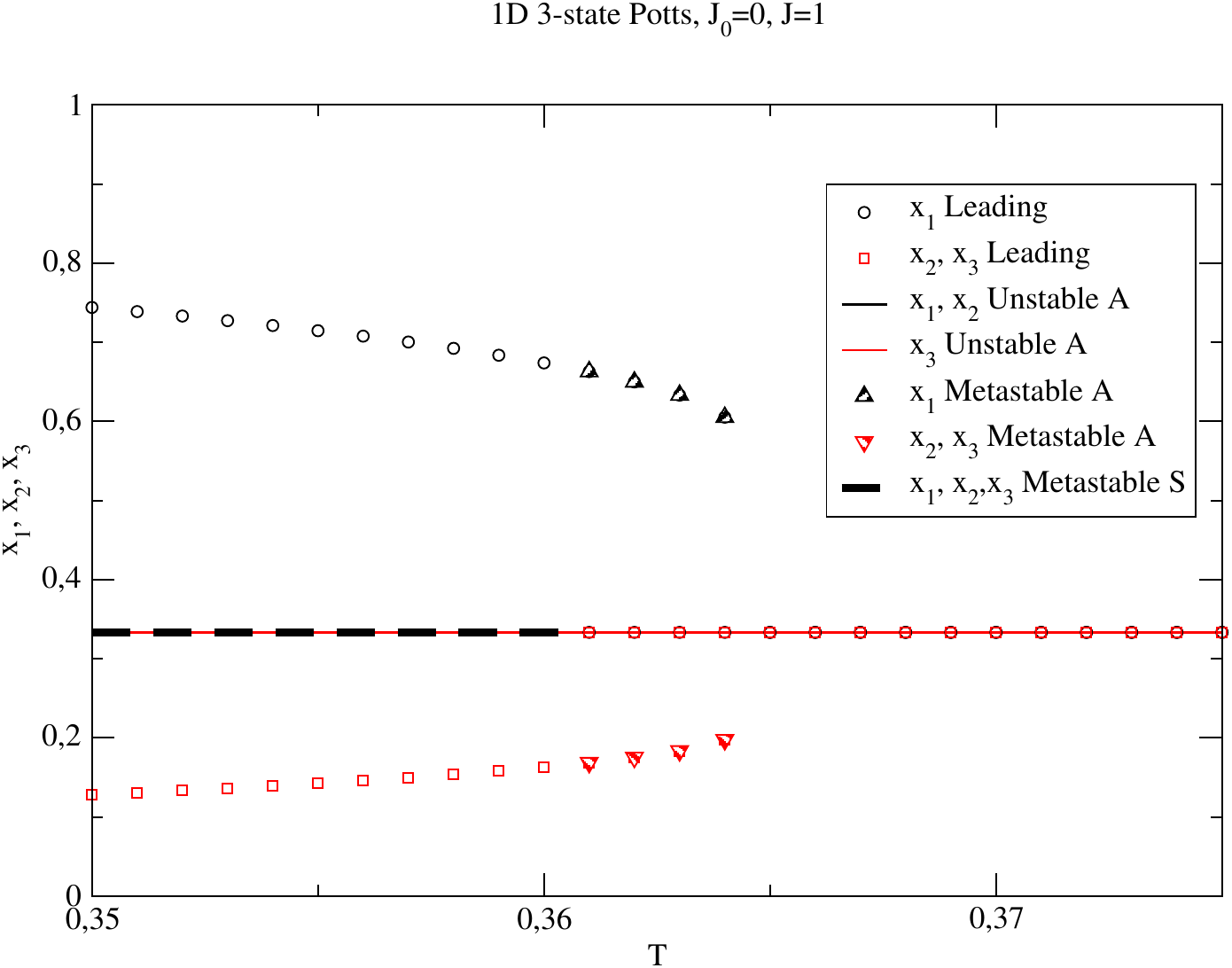}
\caption{(Color online) Particular of Fig.~(\ref{fig1}) near the critical temperature.
}
\label{fig2}
\end{figure}

By plugging Eqs. (\ref{Potts0})-(\ref{Pottsf0}) in Eqs. (\ref{Potts})-(\ref{Pottsf}) we get immediately
the following system of equations and the free energy density:
\begin{eqnarray}
\label{Potts0t}
x_{\sigma}=\frac{e^{\beta J x_\sigma}}{\sum_{\sigma'}e^{\beta J x_{\sigma'}}}, \quad \sigma=1,\ldots,q,
\end{eqnarray}
\begin{eqnarray}
\label{Pottsf0t}
\beta f=-\log\left(\sum_\sigma e^{\beta J x_{\sigma}}\right)+\sum_\sigma \frac{\beta J x_\sigma^2}{2}.
\end{eqnarray}
Eqs. (\ref{Potts0t})-(\ref{Pottsf0t}) give rise to a well known phase transition scenario \cite{Wu}:
a second-order mean-field Ising phase transition sets up only for $q=2$, while for any
$q\geq 3$ there is
a first-order phase transition at the critical value (see Fig.~\ref{fig1}):
\begin{eqnarray}
\label{tc}
\beta_c^{(\mathrm{FO})} J = \frac{2(q-1)}{q-2}\log(q-1).
\end{eqnarray}
It is easy however to see that, besides the first-order transition, there exists also a hidden (unstable) second-order transition
taking place when \cite{mfGlauber}
\begin{eqnarray}
\label{tcso}
\beta_c^{(\mathrm{SO})} J =q.
\end{eqnarray}
At equilibrium the main role of this second-order transition is to determine the temperature $T_c^{(\mathrm{SO})}$ below which
the metastable symmetric state ends to be (locally) stable (see Fig.~\ref{fig2}). We shall see later that this phase transition
scenario holds true (robust) also in the presence of a positive short-range coupling ($H_0\neq 0$).

\section{The three-state 1D case}
We now specialize the above general result to the case in which $H_0$ represents a one-dimensional three-state
Potts Hamiltonian:
\begin{eqnarray}
\label{H0}
H_0(\{\sigma_i\})=-J_0\sum_{i=1}^{N}\delta(\sigma_i,\sigma_{i+1})
\end{eqnarray}
where we have assumed periodic boundary conditions $\sigma_{N+1}=\sigma_1$, and from now on it is understood that
each Potts variable $\sigma$ can take the values 1, 2, and 3.

\subsection{The pure model}
To apply Eqs. (\ref{Potts})-(\ref{Pottsf}) to our case we need to solve the corresponding pure model, which is
a 1D three-state Potts model in the presence of a three-component uniform external field, $\bm{h}$, \textit{i.e.}, we have to calculate the
following partition function
\begin{eqnarray}
\label{Z0}
Z_0(\bm{h})=\sum_{\sigma_1,\ldots,\sigma_N}e^{\beta J_0 \sum_{i=1}^{N}\delta(\sigma_i,\sigma_{i+1})+\beta\sum_{\sigma} h_\sigma N_\sigma}.
\end{eqnarray}
For any finite $N$, we can express (\ref{Z0}) as (``transfer matrix method'')
\begin{eqnarray}
\label{Z01}
Z_0(\bm{h})=\mathrm{Tr}~ \bm{T}^N,
\end{eqnarray}
where $\bm{T}$ is the $3\times 3$ matrix whose elements, $T(\sigma,\sigma')$, for any $\sigma,\sigma'\in\left\{1,2,3\right\}$,
are defined as
\begin{eqnarray}
\label{T}
T(\sigma,\sigma')=\exp\left[\beta J_0\delta(\sigma,\sigma')+\frac{1}{2}\beta h_\sigma+\frac{1}{2}\beta h_{\sigma'}\right].
\end{eqnarray}
For the thermodynamic limit it will be enough to evaluate the eigenvalues of $\bm{T}$, $\lambda_1,~\lambda_2,~\lambda_3$,
the free energy density of the pure model, $f_0$, being given by
\begin{eqnarray}
\label{f0}
-\beta f_0(\beta \bm{h})=\lim_{N\to\infty} \frac{\log\left(Z_0(\bm{h})\right)}{N}=\log(\lambda_{\max}),
\end{eqnarray}
where $\lambda_{\max}$ is such that $|\lambda_{\max}|=\max\left\{|\lambda_1|,~|\lambda_2|,~|\lambda_3|\right\}$.
From Eq. (\ref{T}) we see that the eigenvalues equation reads
\begin{eqnarray}
\label{l0}
\lambda^3+a_2\lambda^2+a_1\lambda+a_0=0,
\end{eqnarray}
where
\begin{eqnarray}
\label{a0}
a_0=\left(3e^{\beta J_0}-e^{3\beta J_0}-2\right)\left(e^{\beta h_1+\beta h_2+\beta h_3}\right),
\end{eqnarray}
\begin{eqnarray}
\label{a1}
a_1=\left(e^{2\beta J_0}-1\right)
\left(e^{\beta h_1+\beta h_2}+e^{\beta h_2+\beta h_3}+e^{\beta h_1+\beta h_3}\right),
\end{eqnarray}
\begin{eqnarray}
\label{a2}
a_2=-e^{\beta J_0}\left(e^{\beta h_1}+e^{\beta h_2}+e^{\beta h_3}\right).
\end{eqnarray}
Eq. (\ref{l0}) is cubic in $\lambda$ so that we can solve it explicitly using the Cardano formula which gives the three roots:
\begin{eqnarray}
\label{l1}
&&\lambda_1=-\frac{1}{3}a_2+(S+T),\\
\label{l2}
&&\lambda_2=-\frac{1}{3}a_2-\frac{1}{2}(S+T)+\frac{\mathrm{i}\sqrt{3}}{2}(S-T),\\
\label{l3}
&&\lambda_3=-\frac{1}{3}a_2-\frac{1}{2}(S+T)-\frac{\mathrm{i}\sqrt{3}}{2}(S-T),
\end{eqnarray}
where
\begin{eqnarray}
\label{ST}
S=\left(R+D^{\frac{1}{2}}\right)^{\frac{1}{3}}, \quad
T=\left(R-D^{\frac{1}{2}}\right)^{\frac{1}{3}},
\end{eqnarray}
\begin{eqnarray}
\label{D}
&& D=Q^3+R^2, \quad
Q=\frac{3a_1-{a_2}^2}{9},\\
\label{R}
&& R=\frac{9a_1a_2-27a_0-{2a_2}^3}{54}.
\end{eqnarray}
Once the eigenvalues have been calculated, the magnetizations, $x_{0;\sigma}$, of the pure model in the thermodynamic limit
can be calculated from Eq. (\ref{f0}) as
\begin{eqnarray}
\label{x0}
x_{0;\sigma}=\frac{1}{\lambda_{\max}}\frac{\partial\lambda_{\max}}{\partial \beta h_\sigma}.
\end{eqnarray}
Notice that $\lambda_1$, $\lambda_2$, and $\lambda_3$ are functions of the vector-field $\bm{h}$.
In the case of equal external fields, $h_1=h_2=h_3$, which in particular includes the case $h_1=h_2=h_3=0$,
due to the fact that $a_0$, $a_1$, and $a_2$ are symmetrical in the fields,
Eq. (\ref{x0}) provides always the symmetric solution $x_{0;\sigma}=1/3$, i.e., as expected, in one dimension there is no
phase transition. Less trivial is to evaluate Eq. (\ref{x0}) for an arbitrary external field $\bm{h}$.
To this aim,
from Eqs. (\ref{a0})-(\ref{R}) we see that 
we need to take into account the
following derivatives, with $\{\sigma,\sigma',\sigma''\}=\{1,2,3\}$
\begin{eqnarray}
\label{Da0}
\frac{\partial a_0}{\partial \beta h_\sigma}=a_0
\end{eqnarray}
\begin{eqnarray}
\label{Da1}
\frac{\partial a_1}{\partial \beta h_\sigma}=
\left(e^{2\beta J_0}-1\right)\left(e^{\beta h_\sigma+\beta h_{\sigma'}}+e^{\beta h_\sigma+\beta h_{\sigma''}}\right),
\end{eqnarray}
\begin{eqnarray}
\label{Da2}
\frac{\partial a_2}{\partial \beta h_\sigma}=-e^{\beta J_0}e^{\beta h_\sigma},
\end{eqnarray}
\begin{eqnarray}
\label{DS}
\frac{\partial S}{\partial \beta h_\sigma}=\frac{1}{3}\left(R+D^{\frac{1}{2}}\right)^{-\frac{2}{3}}
\left(\frac{\partial R}{\partial \beta h_\sigma}+\frac{1}{2D^{\frac{1}{2}}}\frac{\partial D}{\partial \beta h_\sigma}\right),
\end{eqnarray}
\begin{eqnarray}
\label{DT}
\frac{\partial T}{\partial \beta h_\sigma}=\frac{1}{3}\left(R-D^{\frac{1}{2}}\right)^{-\frac{2}{3}}
\left(\frac{\partial R}{\partial \beta h_\sigma}-\frac{1}{2D^{\frac{1}{2}}}\frac{\partial D}{\partial \beta h_\sigma}\right),
\end{eqnarray}
\begin{eqnarray}
\label{DD}
\frac{\partial D}{\partial \beta h_\sigma}=3Q^2\frac{\partial Q}{\partial \beta h_\sigma}+2R\frac{\partial R}{\partial \beta h_\sigma},
\end{eqnarray}
\begin{eqnarray}
\label{DQ}
\frac{\partial Q}{\partial \beta h_\sigma}=\frac{3\frac{\partial a_1}{\partial \beta h_\sigma}-{2a_2}\frac{\partial a2}{\partial \beta h_\sigma}}{9},
\end{eqnarray}
\begin{eqnarray}
\label{DR}
\frac{\partial R}{\partial \beta h_\sigma}=\frac{
9\frac{\partial a_1}{\partial \beta h_\sigma}a_2 +
9\frac{\partial a_2}{\partial \beta h_\sigma}a_1 -
27\frac{\partial a_0}{\partial \beta h_\sigma} -
6a_2^2\frac{\partial a_2}{\partial \beta h_\sigma}
}{54}.
\end{eqnarray}

\subsection{The mean-field model}
By performing the effective substitutions $h_\sigma\to J x_\sigma$
in Eqs. (\ref{f0})-(\ref{DR}), Eqs. (\ref{Potts})-(\ref{Pottsf}) take the form
\begin{eqnarray}
\label{x}
x_{\sigma}=\frac{1}{\lambda_{\max}\left(\left\{\beta J x_\sigma\right\}\right)}
\frac{\partial\lambda_{\max}\left(\left\{\beta h_\sigma'\right\}\right)}{\partial \beta h_\sigma}|_{\left\{h_\sigma'=J x_\sigma'\right\}},
\end{eqnarray}
\begin{eqnarray}
\label{f}
\beta f=\sum_\sigma \frac{\beta J x_\sigma^2}{2}-\log(\lambda_{\max}\left(\left\{\beta J x_\sigma\right\}\right)),
\end{eqnarray}
where we have written the explicit dependence on the arguments $\left\{\beta J x_\sigma\right\}$. For the internal energy
per spin $u$ we have
\begin{eqnarray}
\label{u}
u=f-\frac{1}{\beta}\sum_\sigma x_\sigma\log(x_\sigma),
\end{eqnarray}
where the second term corresponds to the entropy per spin. Particularly simple are the expressions corresponding to the
symmetric solution $\{x_\sigma=1/q\}$. Direct application of the transfer matrix method provides (these formulas are valid for
any $q$)
\begin{eqnarray}
\label{fsymm}
\beta f_{\mathrm{symm}}=-\log\left(e^{\beta J_0}+q-1\right)-\frac{\beta J}{2q},
\end{eqnarray}
\begin{eqnarray}
\label{usymm}
u_{\mathrm{symm}}=f_{\mathrm{symm}}+\frac{1}{\beta}\log(q),
\end{eqnarray}
where $f_{\mathrm{symm}}$ and $u_{\mathrm{symm}}$ stand for free energy and energy (per spin) of the symmetric solution.
Notice that, whereas $f$ is continuous at the critical point of a first-order phase transition, $u$ is not.
Eqs. (\ref{u}) and (\ref{usymm}) can be used for evaluating the latent heat per spin $L$ defined as the difference of the energies of the symmetric solution with the non symmetric
one at the critical point of the first-order phase transition:
\begin{eqnarray}
\label{L}
L=\left(u_{\mathrm{symm}}-u\right)_{|_{T=T_c^{(\mathrm{FO})}}}.
\end{eqnarray}
In the case $J_0=0$ one can shows that $L$ reduces to~\cite{Wu}
$L=\beta J (q-2)^2/(2q(q-1))$.
The latent heat is interesting because it quantifies the amount of energy the system requires to ``transform'' a metastable state
(i.e. subleading) into a leading one along the first-order transition, in close analogy with the change of phase of fluids, like
the gas-liquid transition.

Note that in the present work we are not going to consider an additional external field (see Eq. (\ref{HgPotts})):
according to Eqs. (\ref{Potts})-(\ref{Pottsf}), the role that the external field had on the pure model $H_0$, has been now replaced by
the effective magnetizations $J x_\sigma$ to be found self-consistently by Eqs. (\ref{x}).
We could easily consider the presence of an additional external field $\bm{h}$ by simply performing
the effective substitutions $h_\sigma\to J x_\sigma+h_\sigma$ in Eqs. (\ref{f0})-(\ref{DR}), which
does not change the structure of the self-consistent Eqs. (\ref{x}),
but its numerical detailed analysis goes beyond the aim of the present work.

\section{Numerical analysis and physical interpretation of the self-consistent equations}
In this Section we analyze numerically Eqs. (\ref{x}) and (\ref{f}) and
provide the corresponding physical explanation.
It turns out that $\lambda_{\mathrm{max}}$ coincides always with $\lambda_1$.
We find it convenient to distinguish the cases $J_0\geq 0$ and $J_0<0$ both
for $J>0$. Later on we will consider also the case $J<0$.
As we have seen in the previous Section, the pure model, in one dimension, does
not undergo a spontaneous symmetry breaking.
However, from Eqs. (\ref{x}) and (\ref{f}) we see that,
for any positive value
of $J$, the model governed by $H$ turns out to be a mean-field model so that a phase
transition is always expected. Therefore, in our numerical experiments, it will be enough to keep
the value of the long-range coupling fixed at $J=1$ and obverse what happens by changing
the short-range coupling $J_0$.

In general, the trivial and symmetric solution $x_1=x_2=x_3=1/3$
is stable even below the critical temperature $T_c^{(\mathrm{FO})}$ within a finite range
of temperatures $[T_c^{(\mathrm{SO})},T_c^{(\mathrm{FO})}]$,
though in general is not leading (\textit{i.e.}, it is a metastable state),
and below $T_c^{(\mathrm{SO})}$ becomes unstable.

\subsubsection{The case $J>0$, $J_0\geq 0$}
\begin{figure}
\includegraphics[scale=0.35]{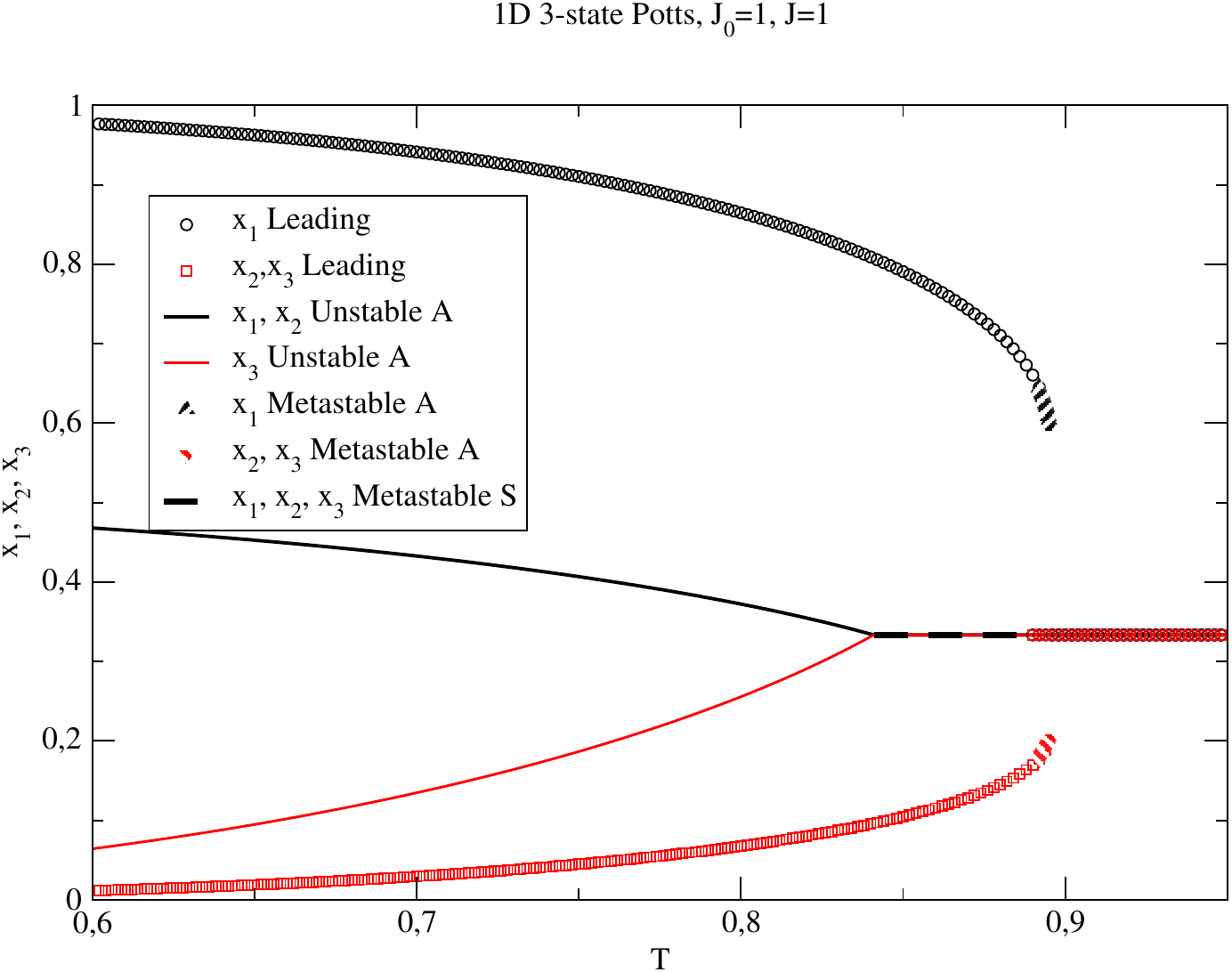}
\caption{(Color online) As in Fig.~(\ref{fig1}) but with $J_0=1$.
}
\label{fig3}
\end{figure}

When we set $J_0\to 0$,
our model coincides with the traditional mean-field model governed
by Eqs. (\ref{Potts0t}) and (\ref{Pottsf0t}).
When $J_0=0$, from Eq. (\ref{tc}) with $q=3$,
we see that for $J=1$ a first-order phase transition develops at the critical
point $T_c^{(\mathrm{FO})}=0.3607$, see Fig.~(\ref{fig1}).
The phase transition is triggered by a broken symmetry mechanism according to which
one of the three components $(1,2,3)$ becomes favored in spite of the other two that remain
equal to each other so that, for $T\to 0$, two components go to 0 and the favored one reaches the value 1.
As expected, when $J_0>0$ we observe a smooth modification of such a scenario,
as reported in Figs.~(\ref{fig2})-(\ref{fig3}). Fig. \ref{fig5} shows that $T_c^{(\mathrm{FO})}$ is an increasing function of $J_0$ for $J_0>0$.
Similarly, Fig. \ref{fig6} shows that the latent heat per particle $L$ is an increasing function of $J_0$ for $J_0>0$.

Besides the first-order phase transition (``dominant'', or ``leading''),
as shown in Figs.~(\ref{fig1}-\ref{fig3}), we observe
the existence of a second-order phase transition which lies
in the sub-space $x_{i_1}=x_{i_2}>x_{i_3}$, where
$i_1,i_2,i_3$, is any permutation of the set of indices $\left\{1,2,3\right\}$
and $x_{i_1}+x_{i_2}+x_{i_3}=1$. This second-order phase transition takes
place at a critical temperature $T_c^{(\mathrm{SO})}$ where the metastable state $x_1=x_2=x_3=1/3$
(\textit{i.e.} a stable state with an higher free energy with respect
to the stable leading state)
becomes unstable and two components become favored against a third one,
so that their values toward $T=0$ are the states (1/2,1/2,0) (and their permutations).
As for any model having a fully connected interaction \ref{HgPotts}, also this second-order transition is mean-field
like with classical critical exponents
(for each order parameter we have $\alpha=0$, $\beta=1/2$, $\gamma=1$, $\delta=1/3$),
as can be checked directly or by applying the general result of Ref.~\cite{MF}.
For $J_0=0$ it is easy to see that the critical temperature of this transition is given by $T_c^{(\mathrm{SO})} =J/q$.
Here the symmetry to be broken seems to be a two-fold one, as can be seen
if we consider that two non zero components are forced to change simultaneously
by the constrain $x_{i_1}=x_{i_2}$.
However, only the state coming from the first-order transition is stable, while
the other turns out to be unstable,
as can be seen from the fact that the initial conditions
giving rise to the second-order phase transition
live in a subspace of $(x_1,x_2,x_3)$ of the kind $x_{i_1}=x_{i_2}$ which has zero volume
in 3 dimensions. A basin of attraction of zero volume corresponds to a unstable state.
In fact, a control of the Hessian of the Landau free energy (\ref{f}) \footnote{\label{NoteL}
The true free energy is given by (\ref{f}) calculated in the solutions of the system (\ref{x}),
while the Landau free energy is represented by (\ref{f}) alone.}
gives, for the solution corresponding to the second-order phase transition,
always one negative eigenvalue.
The presence of a second-order phase transition
in a non disordered three-state mean-field Potts model,
even if hidden in a subspace,
is a quite non obvious and interesting fact: in the subspace $x_{i_1}=x_{i_2}>x_{i_3}$,
as the temperature is decreased,
the system is forced to favor $x_{i_1}$ and $x_{i_2}$ not by a jump, but continuously
with a two-fold broken symmetry mechanism which in turn sets the end of the metastable state.
In the next paragraph we will see that for $J_0<0$ this scenario
is somehow reversed.

\subsubsection{The case $J>0$, $J_0<0$}
\begin{figure}
\includegraphics[scale=0.35]{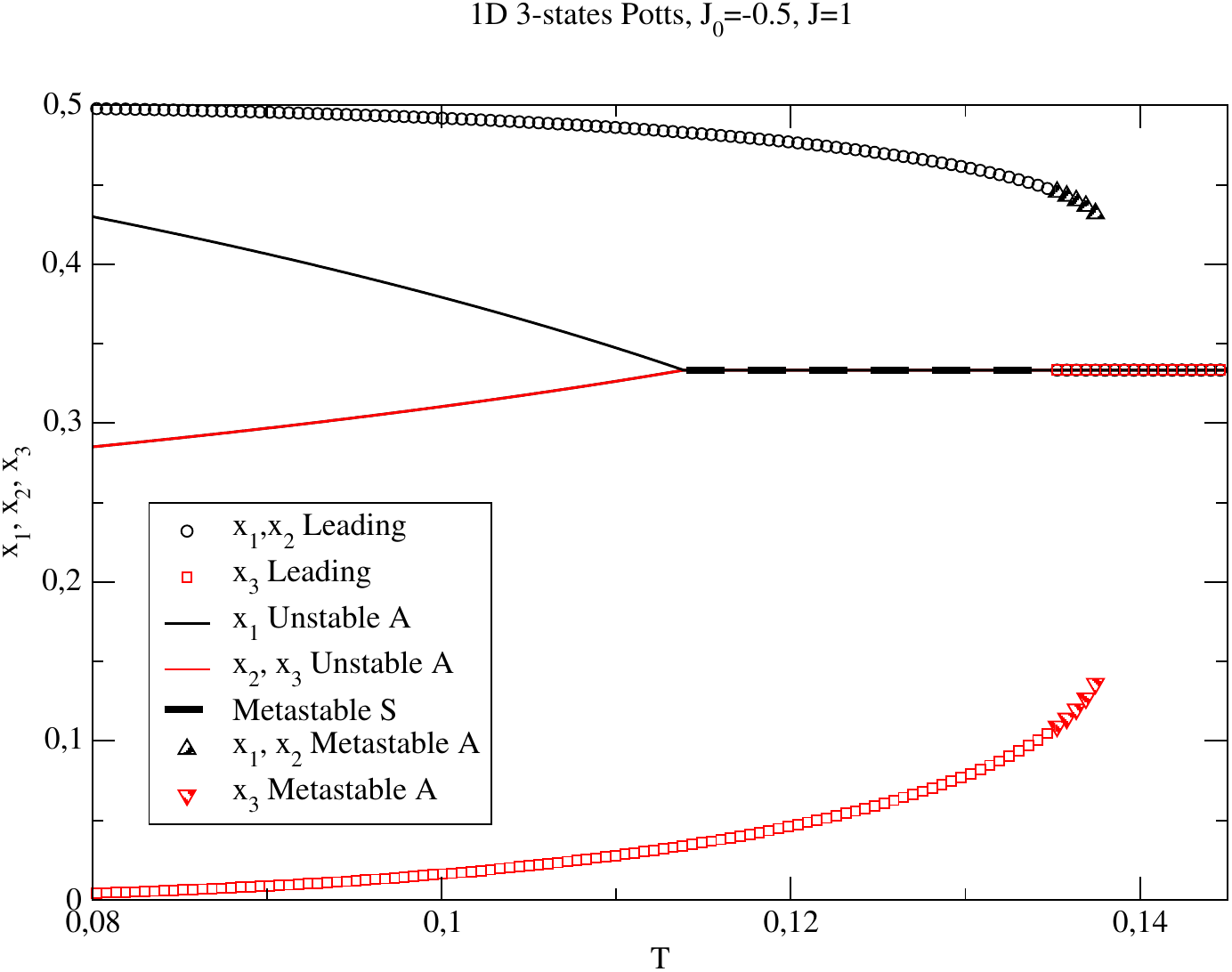}
\caption{(Color online) As in Fig. \ref{fig1} with $J_0=-0.5$.
For $J_0<0$ the Leading states lie on the subspaces $x_{i_1}=x_{i_2}>x_{i_3}$
(in this figure only the case $x_1=x_2>x_3$ is shown),
while the Unstable A states lie on the subspaces $x_{i_1}=x_{i_2}<x_{i_3}$
(in this figure only the case $x_1>x_2=x_3$ is shown), where
$i_1,i_2,i_3$, is any permutation of the set of indices $\left\{1,2,3\right\}$,
and $x_{i_1}+x_{i_2}+x_{i_3}=1$.
Note that, for $J_0<0$, the asymptotic values of the magnetizations toward $T=0$
are $x_{i_1}=x_{i_2}=1/2,~x_{i_3}=0$ and $x_{i_1}=1/2,~x_{i_2}=x_{i_3}=1/4$, for the Leading
and Unstable A states, respectively.
}
\label{fig4}
\end{figure}

When $J_0<0$ the model is still mean field, so that a first-order phase transition
similar to the case $J_0>0$ is also present, but with a corresponding lower value for $T_c^{(\mathrm{FO})}$, as confirmed by Fig.~(\ref{fig4}).
Now, however, due to the fact that $J_0<0$, we must to take into account
that two spins that are consecutive along the 1D chain, at low enough temperatures, cannot have the same value,
so that, among the three components $(1,2,3)$, the favored one(s), if any, along the 1D chain
must be alternated with another one (others).
There are two ways to realize this alternation.
If for example we look for situations in which, at low enough temperatures, the component $1$ is favored,
along the 1D chain we can look for configurations of the kind $(1,2,1,3,1,2,1,3,\ldots)$.
But we can also look for situations in which, for example, both the components $1$ and $2$ are favored
and, at low enough temperatures, the configurations are of the kind $(1,2,1,2,1,2,1,2,\ldots)$.
Notice, for both the situations, with respect to the case $J_0\geq 0$,
the necessary modification of the values $(x_1,x_2,x_3)$ for $T\to 0$ due to the alternations:
now the asymptotic values are either $(1/2,1/4,1/4)$ or $(1/2,1/2,0)$
(and their permutations) for the above former and latter case, respectively.
In both the cases, we have a translational broken symmetry
(similarly to an antiferromagnetic Ising model) but in the latter case,
we have also a further two-fold broken symmetry (since one state, either the state 2 or the state 3, must be excluded).
Interestingly, while the latter phase-transition mechanism corresponds
to a first-order phase transition,
which turns out to be, in shape, quite similar to the phase transition that occurred for $J_0>0$,
the former phase-transition mechanism corresponds to a second-order phase transition.
However, as in the case $J_0\geq 0$,
only the state coming from the first-order transition is stable, while
the other turns out to be unstable,
as seen from the fact that the initial conditions giving rise to the second-order phase transition
live in a subspace of $(x_1,x_2,x_3)$ of the kind $x_{i_1}=x_{i_2}$ which has zero volume
in 3 dimensions or, alternatively, by controlling the Hessian
of the Landau free energy (\ref{f}).
Again, we stress that the presence a second-order phase transition
in a non disordered three-state mean-field Potts model,
even if hidden in a subspace,
is a quite non obvious and interesting fact: in the subspace $x_{i_1}=x_{i_2}<x_{i_3}$,
as the temperature is decreased, the system is forced to favor $x_{i_3}$ not by a jump,
but continuously and, in turn, this transition sets the end of the symmetric metastable state.

\begin{figure}
\includegraphics[scale=0.35]{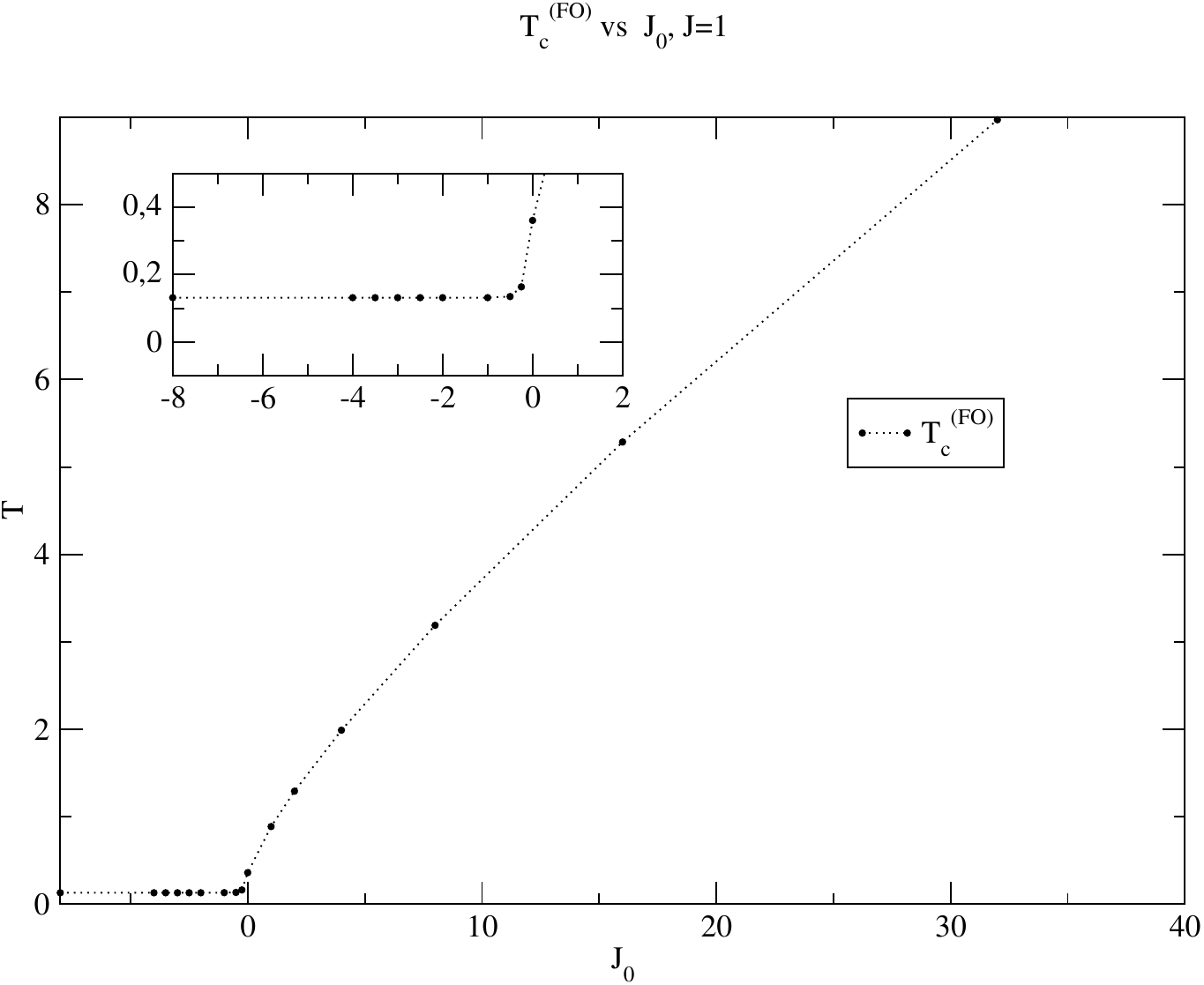}
\caption{(Color online) Behavior of the critical temperature of the first-order
  transition $T_c^{(\mathrm{FO})}$ as a function of $J_0$ with $J=1$ (filled dots; the line is a guide for the eyes). Inset:
  particular in the region of $J_0$ negative.
}
\label{fig5}
\end{figure}

\begin{figure}
\includegraphics[scale=0.35]{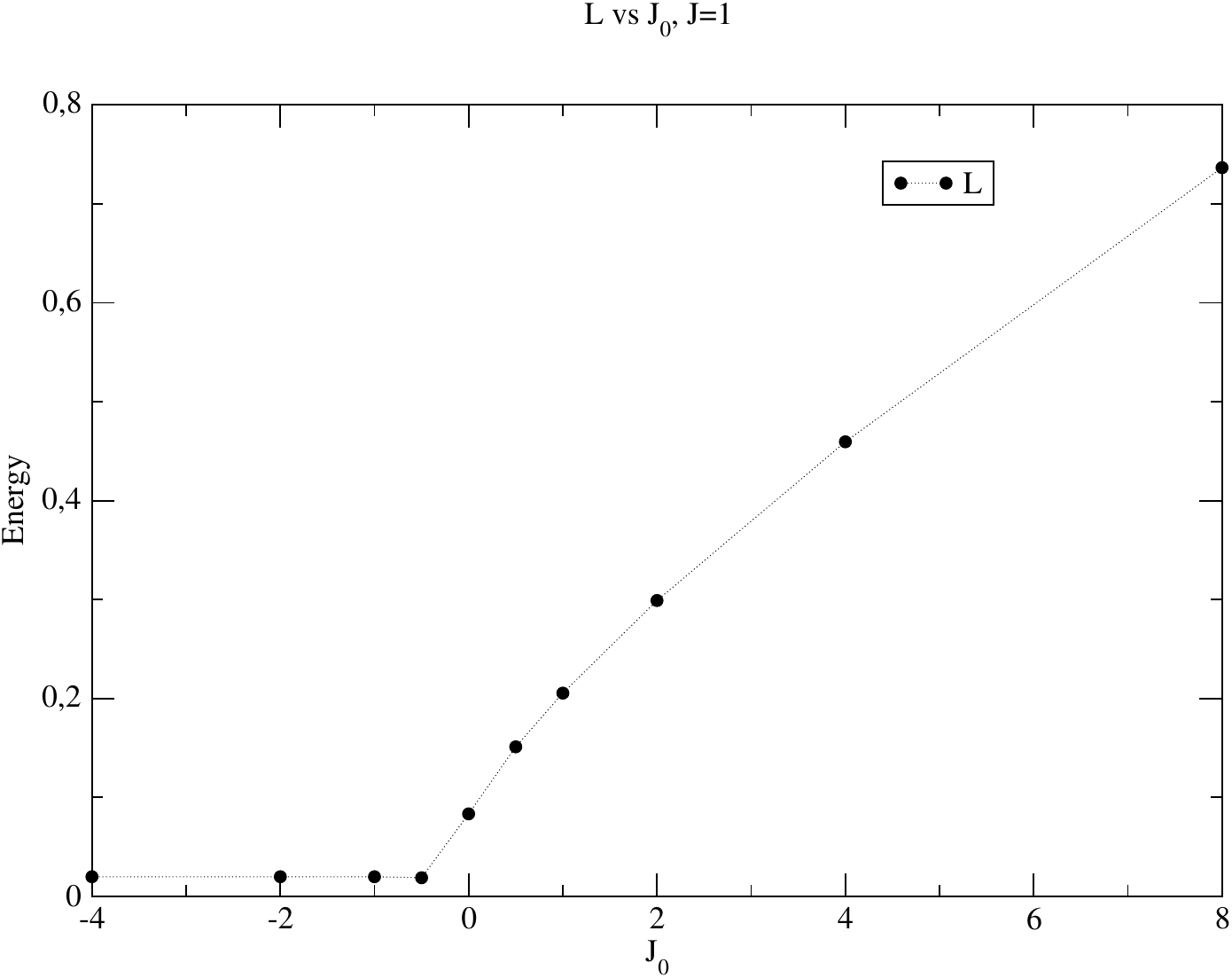}
\caption{(Color online) Behavior of the latent heat per spin $L$ defined in Eq. (\ref{L})
  as a function of $J_0$ with $J=1$ (filled dots; the line is a guide for the eyes).
}
\label{fig6}
\end{figure}

As shown in Fig. \ref{fig5}, the behavior of $T_c^{(\mathrm{FO})}$ as a function of $J_0$ for $J_0<0$ is quite interesting. Differently from the
Ising-like two-state case, where the critical temperature
obeys the equation $\exp(2\beta J_0)\beta J=1$ \cite{SW} (so that it tends to zero for $J_0\to -\infty$), in the present 3-state case
$T_c^{(\mathrm{FO})}$ tends to a finite constant for $J_0\to -\infty$ (see Inset of Fig. \ref{fig5}).
Similarly, as shown in Fig. \ref{fig6}, the latent heat per spin also tends to a finite constant for $J_0\to -\infty$.

\subsubsection{The case $J<0$}
As anticipated in Sec. II,
when $J<0$, the saddle point Eqs. (\ref{x}) are still exact,
as derived from the general theorem presented in \cite{MF}
(while the free energy has a different form with respect to Eq. (\ref{f})).
When $J<0$, Eqs. (\ref{x}) have only the trivial symmetric solution
and no phase transition sets in. A more interesting scenario can emerge under a dynamical approach as done in \cite{mfGlauber} for the
case $J_0=0$ where a dynamical second-order phase transition takes place. The analysis of the dynamical approach for $J_0\neq 0$
will be reported elsewhere.

\section{Correlation functions}
We want to prove now that, as anticipated, the connected correlation function of the
effective mean-field model (\ref{HgPotts}) are not zero and evaluate them in a specific case.
Let us consider for simplicity open boundary conditions and the
two point connected correlation function of two consecutive spins.
It is easy to see that, for the pure $q$-state Potts model at zero external field, we have
\begin{eqnarray}
\label{Corr0}
\mediaT{\delta_{\sigma_i,\sigma_{i+1}}}_0=\frac{e^{\beta J_0}}{e^{\beta J_0}+q-1},
\end{eqnarray}
which implies
\begin{flushleft}
\begin{eqnarray}
\label{Corr1}
C_0(0,\ldots,0)\defi
\sum_\sigma \left(\mediaT{\delta_{\sigma_i,\sigma}\delta_{\sigma,\sigma_{i+1}}}_0
-\mediaT{\delta_{\sigma_i,\sigma}}_0\mediaT{\delta_{\sigma,\sigma_{i+1}}}_0
\right) 
\nonumber \\ 
=\frac{e^{\beta J_0}}{e^{\beta J_0}+q-1}-\frac{1}{q}.~~~~~~~~~~~~~~~~~~~~~~~~~~~~~~~~~~~
  \end{eqnarray}
  \end{flushleft}
Equation (\ref{Corr1}) shows that, in the pure 1D model, the connected correlation functions are zero only in the limit
of infinite temperature and, as expected, they are strictly positive or strictly negative according to the sign of $J_0$, respectively.
Now, on applying Eq. \ref{Corr} to Eq. (\ref{Corr1}), we see that, even above the critical temperature, the connected correlation
functions of the effective mean-field model (\ref{HgPotts}) are not zero.
In fact, for $T>T_c^{(\mathrm{FO})}$, the equilibrium state corresponds to the symmetric solution,
$\{x_i=1/q\}$, which, according to Eq. \ref{Corr}, amounts to have a constant effective external field 
$\bm{h}=\left(h,\ldots,h\right)$ for which
we have trivially $C=C_0(\beta J x_{1},\ldots, \beta J x_{q})=C_0(0,\ldots, 0)$, i.e., as Eq. (\ref{Corr1}). 
For finite size effects see Sec. 3 of Ref. \cite{Twitter}.

\section{Conclusions}
On the base of a general result \cite{MF}, we have considered a three-state Potts model
built over a lattice ring, with coupling $J_0$, and the fully connected graph, with coupling $J$.
This is a non trivial exactly solvable effective mean-field model where new phenomena
emerge as a consequence of the interplay between its finite- and infinite-dimensional character. 
A similar analysis was done in \cite{SW} for the 1D mean-field Ising model (equivalent to a two-state Potts model).
The three-state Potts model, however, shows dramatic differences with respect to the Ising case.
In particular, for given $J>0$, we have found that,
unlike the 1D mean-field Ising model \cite{SW},
the critical temperature $T_c^{(\mathrm{FO})}$ tends to a finite constant when $J_0\to-\infty$.
Similarly, also the latent heat per spin tends to a finite constant for $J_0\to -\infty$.
Moreover, we have found the existence of a hidden continuous phase transition for
both the ferromagnetic, $J_0\geq 0$, and the antiferromagnetic case, $J_0<0$, taking place at
a temperature $T_c^{(\mathrm{SO})}<T_c^{(\mathrm{FO})}$, confirming
the robustness of the scenario found in \cite{mfGlauber} for $J_0=0$ and $J>0$.
However, for $J_0<0$, the system has an antiferromagnetic
feature, the ground state being characterized by a totally different symmetry
with respect to the case $J_0>0$ (compare the asymptotic values toward $T=0$
of Figs. \ref{fig1} and \ref{fig4}).
Concerning the case $J<0$, at equilibrium the only possible stable state is the symmetric one,
while a more interesting scenario emerges under a dynamical approach as done in \cite{mfGlauber} for $J_0=0$,
where the system undergoes only second-order phase transitions (stable).
The extension of the dynamical analysis for $J_0\neq 0$ will be reported elsewhere.
Finally, we have evaluated the connected correlation functions of nearest spins and proven
that they are not zero, even in the paramagnetic phase.

\section*{Acknowledgments}
M. O. acknowledges Grant CNPq 09/2018 - PQ
  (Brazil). F. M. thanks UAEU UPAR Grant No.
  31S391.



\end{document}